# A Survey on Post-Quantum Cryptography: State-of-the-Art and Challenges

Marel Alvarado[1], Luke Gayler[1], Alex Seals[1], Tao Wang[2], and Tao Hou[1]

[1]Texas State University, {marel.alvaradoiii, lgayler, ajs213, taohou}@txstate.edu

[2]University of North Carolina at Charlotte, twang27@charlotte.edu

**Abstract**

The paper explains that post quantum cryptography is necessary due to the introduction of quantum computing causing certain algorithms to be broken. We analyze the different types of post-quantum cryptography, quantum cryptography and quantum-resistant cryptography, to provide a thorough understanding of the current solutions to the problems and their limitations. We explain the current state of quantum computing and how it has changed over time while discussing possible attacks on both types of post-quantum cryptography. Next, current post-quantum algorithms are discussed, and implementations are demonstrated. Lastly, we conclude that due to quantum cryptography's present limitations it is not a viable solution like it is often presented to be and that it is currently better to use quantum-resistant cryptography.

## 1 Introduction

The importance of studying and developing post-quantum cryptography can be understood by first understanding what traditional cryptography is, why we use it, how it works, and its weaknesses. Traditional cryptography has been used for thousands of years [32] to ensure the security of information so that malicious parties cannot tamper with important data. The idea behind traditional cryptography is to scramble or encrypt plaintext messages/information into a cipher-text by using a certain algorithm so that a third party cannot decrypt the message/information. Computers rely on these traditional algorithms so that our information can be secure.

A prominent example of our reliance on these algorithms is our use of Transport Layer Security (TLS) to encrypt our network traffic and ensure no third party can violate the confidentiality, integrity, or authenticity of this traffic. TLS uses a certain algorithm called Rivest-Shamir-Adleman (RSA) encryption which is a type of public-key cryptography. RSA was originally developed in 1978 by Ron Rivest, Adi Shamir, and Leonard Adleman to replace the National Bureau of Standards algorithm [1]. RSA works through using two systems, Public-key encryption and by adding digital signatures to the messages/information. RSA can be explained with the help of the commonly used Alice and Bob example. Let us say Alice wants to send a private message to Bob using the RSA algorithm. The RSA algorithm uses two different keys. The public key, available to everybody and used to encrypt the message, and the private key, only available to the owner (Bob) and used to decrypt the message. To generate the public key, we multiply two large prime numbers p and q together to create n, which is the modulus and choose a value, e, that is relatively prime to (p-1) * (q-1). The modulus n and exponent e together gives us the public key, the public key pair would look like (n,e). Now, to generate the private key, d, we use the formula:

$$ed = 1 mod (p-1) \cdot (q-1)$$

The private key pair would look like (n,d). For RSA to work, the message must be converted to a series of numbers. From here, Alice would generate the cipher-text, c, using:

$$c = S^e mod\, n$$

where S is the plaintext. Then, Bob would use the RSA decryption formula:

$$c = S^e mod\, n$$

to retrieve the plaintext [3]. An additional layer of protection can be added to ensure the message sender is who they say they are. This is achieved in RSA by adding a signature, or unique string of digits generated through a hash function, to the original data, and using same encryption and decryption formulas as before.

While RSA is extremely effective for current computers, its effectiveness relies on the assumption of the difficulty of computing n [3] with our current technology. This reliance on mathematical uncertainties is prevalent within many algorithms used in today's world. However, it turns out that some of these assumptions are only applicable to the current capabilities of computers and are theoretically broken by the creation of quantum computing. RSA specifically will not be secure anymore because of a quantum algorithm known as Shor's algorithm, developed in theory in 1994 by Peter Shor. Shor's algorithm with Beauregard's improvements reduces the complexity of finding a prime number to $O(n^3 \log n)$ and uses 2n + 3 qubits [5], which makes solving the problem take a significantly shorter amount of time. The fact that our data will be compromised by algorithms like Shor's is a leading motivator for research into quantum-safe cryptographic measures.

## 2 Post-Quantum Cryptography Preliminaries

Post-quantum cryptography contains two different approaches, quantum cryptography and quantum-resistant traditional cryptography. The first approach, quantum cryptography was first theorized in 1970 by Stephen Wiesner but was only published by him in 1983 in "Conjugate Coding", where he theorized a quantum communication channel, and a type of quantum money. The first difference between quantum cryptography and traditional cryptography is that quantum uses a unit called qubits instead of bits. A qubit has two distinct states of "0" and "1" much like a bit, but a qubit can also exist in superposition states. These superposition states allow the qubit to exist in both a pure form and a mixed state, which allows many more states to be represented in a smaller amount. By pure form and a mixed state, it means that a qubit can be represented as "1" in different ways. For example, a qubit could be "1" in a ground state or it could be "1" in an excited state. These superpositions provide the qubit with a lot more representation per unit than bits. A single qubit can be denoted as the matrices:

$$|0\rangle = [1\ 0]$$
$$|1\rangle = [0\ 1]$$

Another fundamental difference between the two cryptographies is that quantum cryptography relies on quantum mechanics, a type of theory behind atoms and particle physics, so the properties behind quantum are concrete whereas traditional cryptography relies on mathematical uncertainties. The properties that are currently most important to quantum cryptography are the Heisenberg Uncertainty Principle, Quantum Entanglement, and the No-Cloning Theorem. The Heisenberg Uncertainty principle establishes that due to how some pairs of physical properties relate to one another, it is impossible to measure any object without disturbing it [8]. The Heisenberg Uncertainty can be utilized in cryptographic encryption to detect eavesdropping as any observance of data will cause it to be disturbed. Quantum Entanglement is a feature of qubits in which no matter the distance between two entangled qubits, the measurements will show a high correlation but will not be able to tell what value the qubit has. The use of quantum entanglement is essential for long-distance quantum key distribution [9]. The no-cloning theorem states that you cannot create an identical copy of a quantum state. These three properties are utilized for quantum key distribution (QKD) which is a widely researched type of cryptographic protocol that uses quantum mechanics to produce keys for encrypting and decrypting messages. On top of the quantum-based cryptography measures there are four main types of traditional quantum-resistant cryptography: lattice-based, code-based, hash-based, and multivariate polynomial. While these four types do not rely on quantum mechanics, they are safe from the current quantum algorithms because they do not rely on the hardness of factoring and discrete logarithm problems which have been solved with Shor's algorithm. These traditional post-quantum cryptographic measures are necessary because widespread QKD is not yet feasible due to the inability to transmit qubits across the entire world. Also, the reliability of qubits is not at an adequate level to support main-stream usage. Next, we will discuss the current state of quantum.

## 3 Current Status of Quantum as It Pertains to QKD

In the past 25 years there has been much improvement and development in many parts of quantum computers. First, is the number of qubits available, which has increased from 3-qubits in 1998 [12] to 433-qubits with IBM's Osprey Chip in 2022 [13] with an estimated 1,121-qubit processor estimated to be achieved in 2023. As more qubits are available, researchers are more accurately able to assess the impact of quantum algorithms because certain quantum algorithms require a lot of qubits. Second, is the distance at which we can reliably transmit quantum entangled particles across Earth's surface, which has been demonstrated at 1,120 kilometers (about 695.94 mi) by researchers in China using ground stations that transmit to a satellite and the satellite transmits to the other ground station [9]. This distance is significant progress as the transmission loss across distances has been a barrier to large-scale quantum networks [17]. When long distance transmission of quantum entangled particles is possible, then QKD will become more feasible because systems can be distributed across the world instead of needing to be close in proximity. Lastly, the overall reliability of a quantum computer's qubits is still an issue, but IBM has recently developed error mitigating strategies that reduce the amount of noise by running a quantum computer's circuit multiple times, so the error rate is lower, but this is at the cost of performance [15]. Reliability of a qubit is important in obtaining correct results from quantum algorithms, so with accurate qubits researchers will be able to test theoretical quantum algorithms and ensure the result is accurate. Overall, quantum computing is making slow but steady improvements but again, it is not sufficient for practical use yet. IBM researchers estimate that quantum computing will not be able to solve substantial engineering problems until about 2033 [16]. Next, we will discuss some modern quantum-resistant algorithms, their strengths and weaknesses, and explain and discuss a recent implementation of one of the most recent QKD algorithms developed, Twin fields quantum key distribution (TFQKD).

## 4 Recent Cryptographic Methods

**4.1 CRYSTALS-Kyber:** Chosen by NIST (National Institute for Standards and Technology) in 2022 for general encryption, this public-key encryption method utilizes complicated path-building within lattices to drastically increase computational time for quantum processors (10). To introduce the implementation, for background, the NTT (Number Theoretic Transform) will mathematically transform an input vector into a different vector by using the input vector's values. The XOF (Extendable output function) will generate hashes of desired length; versions of XOF include SHAKE-128 and SHAKE-256. The CBD function will generate noise from an input by using the centered binomial distribution. Additionally, the environment for Kyber will use constant integers k, q, du, and dv.

To generate the keys, first, two values a and b are made from an SHA3 512 hash of a randomized byte array of length 32. Then, in a k * k sized two-dimensional matrix A, for every position in the matrix (i,j), the value at that position becomes an NTT representation of an XOF SHAKE-128 hash generated with parameters a, i, and j. In a one-dimensional matrix of length k, named S, each value becomes a CBD of an XOF SHAKE-256 hash generated with parameters b and the values' literal indexes (for example, if setting the third value in the matrix, the parameters for the hash would be 'b' and 3). Set S to an NTT representation of itself. Another one-dimensional matrix of length k, named E, will be generated the same way as S was generated, but instead of using the values' literal positions as a parameter, the sum of the position and 'k' will be used (for example, if 'k' is 5, and the fourth value of the matrix is being set, then the parameters used would be 'b' and 5 + 4 = 9). Similarly, set E to an NTT representation of itself. Now, a new matrix T will become the sum of E with the matrix product of A and S. The public key will be generated by encoding the concatenation of 'T mod q' with 'a' whilst the private key will be generated by encoding S mod q.

In this environment, we will define another constant integer 'n', which is the product of the length of the private key with '$\frac{96}{k}$'. To encrypt a message 'm', given the receiver's public key 'pk' and a randomized byte array of length 32 'c', first, 'a' is set to "$pk + 12 \cdot k \cdot \frac{n}{8}$". In a k * k sized two-dimensional matrix A, for every position in the matrix (i,j), the value at that position becomes an NTT representation of an XOF SHAKE-128 hash generated with parameters a, i, and j. In a one-dimensional matrix of length k, named 'r', each value becomes a CBD of a SHAKE-256 hash generated with parameters 'c' and the values' literal positions. Another one-dimensional matrix of length k, named 'e1', will be generated the same way as 'r' was generated, but instead of using the values' literal positions as parameters, the sum of the position and 'k' will be used. 'R' will become an NTT representation of 'r', and 'u' will become the sum of 'e1' with the inverse NTT of the matrix product of A and R. Then, 'v' will become the sum of the following: the inverse NTT of the matrix product of 'R' with the public key decoded, 'e2', and 'm' decoded and decompressed. Finally, the encrypted message will be the concatenation of 'u' compressed with parameter du and encoded along with 'v' compressed with parameter dv and encoded.

For a receiver with private key 'sk', given a ciphertext 'c', 'u' will become 'c' decoded with parameter 'du' and decompressed, and 'v' will become '$c + du \cdot k \cdot \frac{n}{8}$' decoded with parameter 'dv' and decompressed. Finally, the message is the private key decoded, matrix multiplied with the NTT representation of 'u', which becomes the parameter of inverse NTT, which is then subtracted from 'v', and finally, compressed and encoded to retrieve message 'm'.

Kyber, in addition to its difficult nature, is incredibly compact in size. Kyber comes in three flavors of increasing difficulty: Kyber-512, Kyber-768, and Kyber-1024. For NIST's level of highest security, level 5, the proposed solution is Kyber-1024, which when benchmarked, used only 4,376 bytes to store the keys, allowing continued security even for lower-end devices(11).

**4.2 Classic McEliece:** Chosen by NIST in 2022 for round four of the Post-Quantum Cryptography Standardization process to continue for consideration for standardization, Classic McEliece is a KEM (Key-Encapsulation Mechanism) based on the original McEliece cryptographic system, which has remained reliable for over 40 years [27,28]. The KEM allows two parties to generate a session key which is used for secure communication. Party A generates the session key and party B generates a public and private key. Party A will encrypt the session key with party B's public key and send the encrypted session key to party B. Party B will decrypt the session key using their private key to retrieve the session key. Now both parties have a session key which can be used to "encapsulate" and "decapsulate" messages in a similar fashion to public-key encryption [29].

In Classic McEliece, for the public and private key, a finite field 'F' of order $2^m$ for natural number m, a random monic irreducible polynomial function 'g(z)' of degree 't' which is an element of 'F', a uniformly random subset of F of distinct elements 'A' of size 'n', and a uniform random n-bit string 's' are held. A two-dimensional matrix 'h' of size t * n, where the value for each element at (i,j) is the product of the j'th element of A raised to the power of 'i - 1' with the inverse of g of the i'th element of A. Consider an m'th dimensional vector space 'f' over a finite field of order 2. Another matrix 'Hˆ' of size '$mt \cdot n$' is generated by using vector space isomorphism between 'F' and 'f' to replace each entry of 'h' with vectors of 'f' arranged in columns. If possible, apply Gaussian elimination to 'Hˆ' for systematic matrix H containing a non-systematic component which is matrix 'T', otherwise repeat from the beginning. The public key will be 'T' and the private key is the set containing 's', g(z), and 'A'.

To generate and encapsulate the session key, a uniform random vector 'e' with Hamming weight 't' which is an element of 'f' is held. The systematic matrix H is reconstructed from public key T. Vector '$C_0$' of length 'm * t' is the product of 'H' with 'e', and vector '$C_1$' is a SHAKE256 hash with parameters 2 and 'e'. Vector C, a ciphertext, becomes a concatenation of $C_0$ and $C_1$ of length '$mt + 256$'. The 256-bit session key K becomes a SHAKE256 hash with parameters 1, 'e', and 'C'.

To decapsulate the key using 'C', 'C' is split into '$C_0$' and '$C_1$'. Vector 'v' becomes '$C_0$' with '$n - mt$' number of zeros appended for ($C_0$, 0, 0, …, 0). Define a Goppa code represented by a set containing 'g(z)' and 'A'. Niederreiter decoding is used to find the codeword 'c'. If 'c' exists, then a vector 'e' becomes 'v + c', but if the hamming weight of 'e' is not equal to 't', '$C_0$' is not the product of 'H' and 'e', then 'e' alternatively becomes 's'. Additionally, a variable 'b' equals 1, but if the SHAKE256 hash with parameters 2 and 'e' does not equal '$C_1$', then 'b' becomes 0 and 'e' becomes 's'. Finally, the SHAKE256 hash on 'b', 'e', and 'C' results in the session key 'K' [30].

Given the long-withstanding nature of the McEliece cryptosystem, this derivation is likely to continue succeeding long term as post-quantum cryptography develops. Additionally, this mechanism uses low amounts of computational resources, where encapsulation and decapsulation use 500,000 cycles or less, while CRYSTALS-Kyber peaked at 1,009,448 cycles on Kyber1024-90s, the 90's variant of their most intensive flavor [33, 11]. Unlike kyber, this mechanism trades off storage for performance; the size of the public key generated peaks at 1.4 megabytes, while the CRYSTALS-Kyber public key peaks at 1,568 bytes [30, 11]. While Kyber takes more computational time, in terms of compatibility, Kyber is a very effective choice compared to Classic McEliece.

**4.3 SPHINCS+:** Chosen by NIST in 2022 for digital signing, SPHINCS+ is a stateless hash-based signature scheme reliant on the use of hash functions to allow authentication for a limited number of messages [27]. For background, SPHINCS+ requires use of W-OTS+ (Winternitz One-Time Signature) and XMSS (exTended Merkle Signature Scheme) to build a hypertree, and FORS (Forest of Random Subsets) to build the FORS-hypertree combination.

W-OTS+: Given a SPHINCS+ signing key 'sk', 'w' number of bits in a window at a time, '$W = 2^w$', we generate a public-private vector key pair. A private vector key 'SK' contains values generated by the PRF (pseudo-random function), which generates random sequences of values given a seed key. Here, each value in 'SK' is the PRF of 'sk'. Additionally, the corresponding public vector key 'PK' of the same length as 'SK' consists of values generated by a chaining pseudo-random function on each corresponding value within 'SK' (the chaining pseudo-random function repeatedly applies any hash function such as SHA2-256 a specified number of times on a value). Now with 'SK', a signature for a message can be generated. A given message 'msg' is split into window chunks of 'w' number of bits. Consider the sum '$\sum_i (W - 1 - w_i)$' for every window bit chunks $w_i$, and again split this sum into 'w' number of bits. In a new vector 'σ' of the size of the number of window bit chunks, for every value in 'σ', the i'th value in 'σ' is the result of the chaining pseudo-random function on the i'th value in 'sk', parameterized by the i'th window bit chunk. Now 'σ' becomes the signature for 'msg'. Now with 'σ' and 'msg', the same public key 'PK' can be extracted. Similarly, 'msg' is split into window chunks of 'w' number of bits. Consider the sum '$\sum_i (W - 1 - w_i)$' for every window bit chunks $w_i$, and again split this sum into 'w' number of bits. In a new vector 'pk' of the size of the number of window bit chunks, for every value in 'pk', the i'th value in 'pk' is the result of the chaining pseudo-random function on the i'th value in 'σ', parameterized by 'W - 1 - $w_i$', where $w_i$ is the i'th window bit chunk. Now 'pk' should be the same public vector key 'PK'.

XMSS: Given 'h' tree height and SPHINCS+ signing key 'sk', $2^h$ number of W-OTS+ public-private vector key pairs are generated. A vector containing the private vector keys from all the pairs becomes the XMSS signing key 'SK'. Additionally, each corresponding public vector key is applied to any given hash function, such as SHA2-256, in the same ordering as the corresponding private vector keys in 'SK'. A binary hash tree is created from these hashed public vector keys, where two keys are hashed together to a separate key for all keys to form a binary tree-like structure. For example, if 'h = 3', then $2^3$ = 8 nodes are created. The first layer will contain four nodes. Two of them are hashed to another node and the other two hashed to a separate node. The two nodes that were hashed to are now hashed to a different node. Now that this different node remains without enough other nodes to continue building the binary hash tree, this node becomes the root of the tree. Now with the binary hash tree of public vector keys, this becomes the public key 'PK'. Now an "authentication path" within the binary hash tree can be signed using 'SK'. For a public vector key node within 'PK', use its corresponding private vector key within 'SK' to produce 'σ', the W-OTS+ signature for the message. Additionally, compute 'auth', the authentication path for the index of this public vector key node within 'PK'. The signature for the message becomes '(σ, auth)'. Like W-OTS+, given the signature and the known message 'msg', the XMSS public key can be extracted. Using W-OTS+ the corresponding W-OTS+ public vector key 'pk' is extracted using 'σ' and 'msg'. Using XMSS, with 'pk' hashed by the same hash function used to generate the original XMSS public key, along with 'auth', the original XMSS public key 'PK' is extracted.

Hypertree: Like XMSS, the hypertree for SPHINCS+ will be a tree composed of XMSS key pairs where nodes are "stacked" and the XMSS node above signs the node below. The tree is parameterized by height 'H', total hypertree heigh 'h', and 'd' layers from 0 to 'd - 1', where the 0'th node is the bottommost node and the node at 'd - 1' is the top-most node. Given the SPHINCS+ signing key 'sk', the hyptertree signing key 'SK' becomes 'sk', and the hypertree public key 'PK' becomes the public of the XMSS public key at layer 'd - 1'. In addition to the keys, note that due to the nature of the hypertree, given the tree index $T_i$ of a node in the hypertree where $0 \leq i < d - 1$, the tree index of the next node $T_{i+1}$ and the index of the W-OTS+ public vector key $\lambda_{i+1}$ within that next node (recall it is a binary hash tree) can be found. From $T_i$, $T_{i+1}$ is the '$h - H(i + 1)$' number of most significant bits of $T_i$, and $\lambda_{i+1}$ is the 'H' number of least significant bits of $T_i$. With a way to access every node in the hypertree, the tree can be signed. Given an n-byte digest 'r' in hyper-leaf index $1 \leq \lambda \leq 2^h$, suppose $T_{-1} = \lambda$ and 'σ' is an empty vector. For integer 'i' from 0 to 'd - 1', repeat the following three steps: (1) derive $T_i$ and $\lambda_i$ from $T_{i-1}$, (2) generate XMSS keys '$sk_i$' and '$pk_i$' at the corresponding address of $T_i$, (3) sign 'r' with '$sk_i$' using leaf index $\lambda_i$ to produce $\sigma_i$, append $\sigma_i$ to σ, and set 'r' to '$pk_i$'. The signature of the hypertree is the resulting vector 'σ'. Observing this signing process, 'r' acts as a temporary holder for the public key of the previous node, then the current iteration signs 'r' using the current node's signing key, hence a "linking" between nodes to form a greater authentication path. The hypertree signing key can now be used to extract the public key from the hypertree. Given the signing key 'σ', for integer 'i' from 0 to 'd - 1', repeat the following three steps: (1) derive $T_i$ and $\lambda_i$ from $T_{i-1}$, (2) extract the current XMSS node's public key '$pk_i$' from '$\sigma_i$' by using the message at the corresponding address of '$T_i$' and the leaf index '$\lambda_i$', (3) set 'r' to '$pk_i$'. Once 'r' is updated with the final public key, its value will be $pk_{d-1}$, the XMSS public key at layer 'd - 1', which is the hypertree public key 'PK'.

FORS: Given a SPHINCS+ signing key 'sk', 'address 'ADRS', 'k' number of FORS trees, and 't = $2^a$' number of tree leaves for tree height 'a', for every FORS tree, the signing key '$SK_i$' for the i'th tree becomes a vector of length 't', where the j'th value in the vector is the result of the pseudorandom function applied to 'sk' with parameter 'ADRS'. Subsequently for the public key, '$PK_i$', apply any hash function to every value in '$SK_i$' and create a binary hash tree from the values in '$SK_i$' in the same manner as creating the public binary hash tree key in XMSS. The resulting hash tree is the public key '$PK_i$'. Finally, overall, for FORS, the signing key 'SK' is the collection of the signing keys from all the FORS trees, and the public key 'PK' is the result of any hash function applied to the collection of all the FORS trees' public keys with parameter 'ADRS'. These keys can be used to sign an m-bit digest 'md'. 'md' is split into chunks ($m_1$, …, $m_k$) of 'a' number of bits. For every chunk, for $m_i$, retrieve the leaf at $m_i$ in the i'th FORS subtree by retrieving the i'th signing key in 'SK' and retrieving $s_{m_i}$ from the i'th signing key. Apply the same hash function used when generating the binary hash tree public keys for the FORS subtrees to $s_{m_i}$ to generate '$auth_i$' authentication path. Finally, the signing vector becomes a collection of collections σ = ( ($s_{m_1}$, $auth_1$), …, ($s_{m_k}$, $auth_k$) ). Now σ can be used with 'md' to extract the public key. 'md' is split into chunks ($m_1$, …, $m_k$) of 'a' number of bits. For every chunk, for $m_i$, recompute the public key '$PK_i$' of the i'th FORS subtree by using the same hash function used during FORS signing applied to $s_i$, the first value in the i'th collection from σ. Subsequently, on a collection of these FORS subtrees public keys generated, apply the same hash function used when generating the FORS 'PK' key, with parameter ADRS, to reproduce 'PK', the original FORS public key.

SPHINCS+ signing: Given message 'msg', $pk_1$ public key of a hypertree, and SPHINCS+ signing keys $sk_1$ and $sk_2$, generate 'R' as the result of applying the pseudorandom function to 'msg' with parameter $sk_2$. Generate 'md' and 'ADRS' resulting from any hash function applied to 'msg' with parameters '$pk_1$' and 'R'. Create signature '$σ^f$' by signing 'md' with FORS at address 'ADRS' using $sk_1$, and let 'PK' be the corresponding FORS public key. Sign 'PK' with the hypertree for '$σ^{ht}$'. Finally, the signature for the message is the collection (R, $σ^f$, $σ^{ht}$).

SPHINCS+ verification: Given $pk_1$ public key of the hypertree from signing, SPHINCS+ signature (R, $σ^f$, $σ^{ht}$), and message 'msg', 'md' and 'ADRS' are produced by applying the same hash function (used to generate 'md' and 'ADRS' from signing) to 'msg' with parameters '$pk_1$' and 'R'. Extract '$PK^f$' using 'md and 'σ$^f$' to get the public key of the FORS at 'ADRS'. Extract 'PK$^{ht}$' by using 'PK$^f$' and 'σ$^{ht}$' to get the public key of the hypertree at the leaf index from 'ADRS'. If 'PK$^{ht}$' is the same as 'pk$_1$', the message is authenticated, otherwise the message is not authenticated [31, 27].

In terms of security, SPHINCS+ is made increasingly secure by the assumed hash functions used. SPHINCS+ does not use set defined hash functions, but instead the implementers decide what hash function to use, so if strong hash functions are used when implementing SPHINCS+, this becomes a very strong signature system. Additionally, most implementations of SPHINCS+ lead to high computational times and concerning signature size; on optimized implementations supporting the AVX2 instruction set, SPHINCS+ with SHA2 hashing used, at minimum, 33.7 million cycles to generate the message signature, which outdoes McEliece and Kyber by a very high margin. Additionally, the signature size for SPHINCS+ peaked at 48,856 bytes, which is much less compared to McEliece, yet Kyber leads by a higher margin. While bulkier, NIST noted, among its four post-quantum choices, that SPHINCS+ is useful as an alternative to its efficient lattice-based competitors for its alternative math approach with hash functions [10, 35].

**4.4 Twin fields quantum key distribution:** Twin fields quantum key distribution (TFQKD) was first proposed in 2018 by Marco Lucamarini et al. [16] and has quickly risen to become a very popular and promising solution to quantum key distribution across long distances. The authors' goal behind their paper was to propose a type of QKD that does not rely on the use of quantum repeaters to transmit across long distances. The motivation to bypass quantum repeaters is that quantum repeaters are not yet a feasible piece of technology despite the recent advancements, as we need something that can be implemented feasibly today [16]. To explain the TFQKD protocol, we will use an Alice and Bob example. First, Alice and Bob will both have a light source and an interferometer while a third-party Charlie will have a beam splitter and detectors. Next, these light sources will be used to create two random phases between $[0,2\pi)$ and will then be encoded with two sets of secret bits and bases, which are essentially encoding phrases. These are sent as pulses to a third-party station Charlie, who could even be a malicious individual. Charlie will then utilize a beam splitter to overlap the pulses to measure them. Charlie will then tell Alice and Bob which of the detectors lit up, one lights up if the bits are equal while the other lights up if they are different. Due to having the detectors, Charlie will also be able to tell whether the bits were equal or not but not what their values are, thus keeping the protection against eavesdropping intact. After Alice and Bob know which detector lights up, they will publicly announce the random phase and the encoding phrase. If their phases are equal, they will be considered "twins", and these "twins" will be saved while the phases that are not equal will be discarded. Additionally, to protect against eavesdropping, decoy states are employed, which are fake or decoy quantum states that are purely used to detect eavesdropping. The phase "randomization" mentioned earlier and decoy states are necessary to support long-distance TFQKD's security. It is important to note that there are performance impacts to adding decoy states, so to reduce this the devices utilizing TFQKD must alternate between code and decoy modes.

A recent implementation of TFQKD was demonstrated in 2022 by a group of researchers across an 830-km fiber cable. To achieve this feat, the researchers used a laser with a central wavelength of 1,550.12 nm and a linewidth of 0.1 kHz, which was then split in two and sent to both ends (Alice and Bob for example), across 411.55 km (about 255.73 mi) each through servo channels to their source parts. The two light beams were again split in half at the source part using beam splitters to allow one to be used for locking each of their local laser's central wavelength and the other to get sent through a chopper, encoder, and regulator to ultimately get sent into the quantum channel. At the source part, their light beams were stabilized to a fixed state of polarization by a polarization compensation module (PCM). The PCM then interfered with the light beam which was detected by two positive-intrinsic-negative (PIN) detectors. Then, using the values detected by the PINs, a homodyne optical phase-locked loop (OPLL) stabilized the difference between the values and made their difference near zero and had its noise reduced using a negative-feedback phase modulator (PM). Next, the other half of the original light beam, which was now locked at the correct wavelength, was passed through the chopper which consisted of two acousto-optic modulators (AOMs) operating at an 80-MHz frequency shift and one intensity modulator (IM) modulating at a width of 60 picoseconds at 4Ghz. Inside of the chopper, the IM modulated the light into a pulse train. The AOM+ and AOM- components then chopped up the pulse train into a time-multiplexed reference part and a quantum part. Next, the quantum part was operated on in the encoder which consisted of three IMs and two PMs. The encoder alternated code and decoy modes like we mentioned earlier. While the encoder was in code mode, the IMs created one quantum state with an intensity, and the PMs generated random phases between and $\left[0, \frac{\pi}{2}\right]$. While in decoy mode, the IMs created three separate quantum states with separate intensities while the PMs created random phases between $[0, 2\pi]$. Next, the light went into the regulator to get high interference. The high interference was achieved by using a dispersion compensation module (DCM) using fiber-Bragg-grating technology, a variable optical delay (VOD), and a variable optical attenuator (VOA). Lastly, the modified light beams were sent through a quantum channel made of G.654.E ULL optical fiber to a third party's (Charlie for example) beam splitter, then two detectors comprised of NbN thin films checked the lights truth values [19].

# 5 Quantum Attacks on Modern Cryptography

Only a decade after quantum computers were first proposed, scientists and mathematicians were already developing algorithms to break common cryptography methods. In 1994, Daniel Simon developed Simon's algorithm. Then, in the same year, mathematician Peter Shor developed an algorithm based on modular arithmetic that would be able to break RSA and Diffie–Hellman key encryption. Two years later in 1996, computer scientist Lov Grover devised an algorithm that highlights the speed-up that could be achieved with quantum computers.

When it comes to breaking the encryption standards we use today, Shor's algorithm is seen as the biggest threat. The security of RSA encryption comes from the difficulty of finding the prime factors of very large numbers. This task is the very thing Shor's algorithm was designed to do. It works by performing calculations that bring it closer to a solution with each step. If we have a number, N, that fits the parameters of RSA encryption, we first choose some guess, a, such that 1 < a < N. Then, we can perform a series of calculations of $a^r \, mod \, N$, where r is increased by 1 for each iteration. Eventually, the results of this equation will begin repeating. The number of values in one repetition is called the order, denoted by r. This step is the order finding sub-routine and it is the quantum portion of the algorithm. It involves modular exponentiation and an inverse quantum Fourier-transform. Once r is found, it can be plugged into the equation $\gcd(a^{\frac{r}{2}} \pm 1, \, N)$. The results of this will be the prime factors of N.

Shor's algorithm is quite intensive when it comes to the required resources. As the value of N increases, the number qubits and gates needed quickly exceeds what we have available. As of now, quantum computing technology has not been able to perform these demonstrations without using some pre-determined parameters to reduce the resource requirements of the system [24]. In 2001, a group of researchers at IBM successfully used Shor's algorithm to find the prime factors of 15. This was done using a process called nuclear magnetic resonance [20]. In 2019, the numbers 15, 21, and 35 were factored using a version of Shor's algorithm on a 6-bit IBM quantum processor [22]. This demonstration used one qubit in what is called the control register. This meant the qubit had to be recycled for each measurement.

Two years later in 2021, an IBM quantum processor was used again to find the prime factors of 21. The number of qubits in this demonstration was reduced to 5. The circuit contained 2 qubits in the work register and 3 qubits in the control register. A sub-processor configuration of IBM's 7 qubit, ibmq-casablanca, and 21 qubit, ibmq-toronto were used. Because of the high resource of Shor's algorithm, the researchers needed to implement a pentagonal circuit mapping that was not available with any single quantum processor. The pentagonal circuit arrangement allowed for more connections between qubits which helped to reduce the number of gates needed to perform the algorithm. In the demonstration, the initial guess was chosen to be 4. For the quantum part of the algorithm, measurements showed probability peaks at 3 and 5. Then using a classical algorithm to perform further calculations, the order was found to be 3. Plugging this into the final step, $\gcd(4^{\frac{3}{2}} \pm 1, \, 21)$, gives 3 and 7 which are the prime factors of 21.

# 6 Proposed Attacks on Quantum Key Distribution

As discussed previously, quantum computers will not only be a powerful tool for breaking encryption. Some of the same principles that make quantum computers so dangerous also allow them to implement powerful security measures as with Quantum Key Distribution. Theoretically, QKD methods should be unbreakable. However, there are imperfections in the generation and measurement of photons, leaving space for an attack to be carried out [18].

One such attack was proposed in 2018 against BB84 QKD protocols [18]. It is known as the photon number splitting attack. A basic requirement of BB84 is sending a photon between two parties, Alice and Bob. Typically, if this photon was intercepted and measured by a third eavesdropping party, their presence would be easily recognizable due to the no-cloning theory. However, generating a single photon can be very difficult. In many cases, when Bob generates a photon, multiple identical photons are generated and sent to Alice. This would allow someone to capture and measure one of the "extra" photons without giving any sign that they are present.

In 2015, four attacks against QKD protocols were proposed [19]. The first, a Man in the Middle Attack (MITM), relies on Eve setting up her own equipment for measurement and generation in the communication channel. She would also need to have access to quantum memory. This MITM attack requires Eve to impersonate both Alice and Bob. In one QKD protocol, an initial set of messages is sent back and forth between Alice and Bob to form a key. The first message is generated by Alice using a particular basis state. Once Bob receives this message, he will need to choose a basis state for measurement. If he chooses the same state as Alice, the message is kept and used as part of the key. If the state does not match, the message is discarded. This is repeated until Alice and Bob have enough matching messages to form a key. If Eve is able to intercept these messages, she could effectively impersonate Alice and Bob. She would intercept Alice's first message and store it in her quantum memory. She could then use the next message sent by Alice to compare with her memory and find collisions in the hashing. This would lead to Eve obtaining an identical copy of Alice's raw key. Eventually, if Eve uses the same protocol as Alice and Bob, all three parties will end up generating the same "private" key. If this key is then used as part of the generation of subsequent keys, then it will be even easier for Eve to obtain these keys.

# 7 Conclusion

Based on our research, quantum cryptography is an emerging field of study with great potential. However, it is not currently an effective countermeasure to quantum computing's nullification of some traditional cryptography algorithms. Namely, there needs to be more research and development into long-distance quantum transmission because currently these systems cannot work across far distances on terrestrial Earth. It is likely that quantum cryptography will be the superior method of resisting attacks in the future, but it is currently quantum-resistant traditional cryptography that is the best method to use. This was concluded because quantum cryptography is in its infancy and has many fundamental issues that likely will not be solved for a long time, and with quantum computing approaching at a rapid pace, we need to be able to protect our data in the meantime, which quantum-resistant traditional cryptography is capable of. On top of the fundamental issues, quantum cryptography requires highly specialized equipment to operate while quantum-resistant traditional cryptography operates on current computers, so there is not a high cost associated with shifting to quantum-resistant as the standard. Even though quantum-resistant traditional cryptography is already functional, further research needs to be done to optimize and develop new algorithms. By developing attacks against them, researchers can modify the algorithms to ensure maximum security.